# Measurements of the $Q^2$-Dependence of the Proton and Deuteron Spin Structure Functions $g_1^p$ and $g_1^d$

**The E143 Collaboration***

K. Abe,[15] T. Akagi,[12,15] P. L. Anthony,[12] R. Antonov,[11] R. G. Arnold,[1] T. Averett,[16] H. R. Band,[17] J. M. Bauer,[7] H. Borel,[5] P. E. Bosted,[1] V. Breton,[3] J. Button-Shafer,[7] J. P. Chen,[16] T. E. Chupp,[8] J. Clendenin,[12] C. Comptour,[3] K. P. Coulter,[8] G. Court,[12,*] D. Crabb,[16] M. Daoudi,[12] D. Day,[16] F. S. Dietrich,[6] J. Dunne,[1] H. Dutz,[12,**] R. Erbacher,[12,13] J. Fellbaum,[1] A. Feltham,[2] H. Fonvieille,[3] E. Frlez,[16] D. Garvey,[9] R. Gearhart,[12] J. Gomez,[4] P. Grenier,[5] K. A. Griffioen,[11,†] S. Hoibraten,[16,§] E. W. Hughes,[12] C. Hyde-Wright,[10] J. R. Johnson,[17] D. Kawall,[13] A. Klein,[10] S. E. Kuhn,[10] M. Kuriki,[15] R. Lindgren,[16] T. J. Liu,[16] R. M. Lombard-Nelsen,[5] J. Marroncle,[5] T. Maruyama,[12] X. K. Maruyama,[9] J. McCarthy,[16] W. Meyer,[12,**] Z.-E. Meziani,[13,14] R. Minehart,[16] J. Mitchell,[4] J. Morgenstern,[5] G. G. Petratos,[12,‡] R. Pitthan,[12] D. Pocanic,[16] C. Prescott,[12] R. Prepost,[17] P. Raines,[11] B. Raue,[10] D. Reyna,[1] A. Rijllart,[12,††] Y. Roblin,[3] L. S. Rochester,[12] S. E. Rock,[1] O. A. Rondon,[16] I. Sick,[2] L. C. Smith,[16] T. B. Smith,[8] M. Spengos,[1] F. Staley,[5] P. Steiner,[2] S. St.Lorant,[12] L. M. Stuart,[12] F. Suekane,[15] Z. M. Szalata,[1] H. Tang,[12] Y. Terrien,[5] T. Usher,[12] D. Walz,[12] J. L. White,[1] K. Witte,[12] C. C. Young,[12] B. Youngman,[12] H. Yuta,[15] G. Zapalac,[17] B. Zihlmann,[2] D. Zimmermann[16]

[1]The American University, Washington, D.C. 20016
[2]Institut für Physik der Universität Basel, CH–4056 Basel, Switzerland
[3]LPC IN2P3/CNRS, University Blaise Pascal, F–63170 Aubiere Cedex, France
[4]CEBAF, Newport News, Virginia 23606
[5]DAPNIA-Service de Physique Nucleaire Centre d'Etudes de Saclay, F–91191 Gif/Yvette, France
[6]Lawrence Livermore National Laboratory, Livermore, California 94550
[7]University of Massachusetts, Amherst, Massachusetts 01003
[8]University of Michigan, Ann Arbor, Michigan 48109
[9]Naval Postgraduate School, Monterey, California 93943
[10]Old Dominion University, Norfolk, Virginia 23529
[11]University of Pennsylvania, Philadelphia, Pennsylvania 19104
[12]Stanford Linear Accelerator Center, Stanford, California 94309
[13]Stanford University, Stanford, California 94305
[14]Temple University, Philadelphia, Pennsylvania 19122
[15]Tohoku University, Sendai 980, Japan
[16]University of Virginia, Charlottesville, Virginia 22901
[17]University of Wisconsin, Madison, Wisconsin 53706

*Submitted to Physics Letters B*

*Work supported in part by Department of Energy contract DE–AC03–76SF00515 (SLAC), and by other Department of Energy contracts for CEBAF, LLNL, ODU, Stanford, Virginia and Wisconsin; by the National Science Foundation (American, Massachusetts, Michigan, ODU, and U. Penn.); by the Schweizersche Nationalfonds (Basel); by the Commonwealth of Virginia (Virginia); by the Centre National de la Recherche Scientifique and the Commissariat a l'Energie Atomique (French groups); and by the Japanese Ministry of Education, Science, and Culture (Tohoku).

*Permanent address: Oliver Lodge Lab, University of Liverpool, Liverpool, U.K.
**Permanent address: University of Bonn, D-53113 Bonn, Germany.
†Present address: College of William and Mary, Williamsburg, Virginia 23187.
§Permanent address: FFIYM, P.O. Box 25, N-2007 Kjeller, Norway.
‡Present address: Kent State University, Kent, Ohio 44242.
††Permanent address: CERN, 1211 Geneva 23, Switzerland.

## ABSTRACT

The ratio $g_1/F_1$ has been measured over the range $0.03 < x < 0.6$ and $0.3 < Q^2 < 10$ (GeV/c)$^2$ using deep-inelastic scattering of polarized electrons from polarized protons and deuterons. We find $g_1/F_1$ to be consistent with no $Q^2$-dependence at fixed $x$ in the deep-inelastic region $Q^2 > 1$ (GeV/c)$^2$. A trend is observed for $g_1/F_1$ to decrease at lower $Q^2$. Fits to world data with and without a possible $Q^2$-dependence in $g_1/F_1$ are in agreement with the Bjorken sum rule, but $\Delta q$ is substantially less than the quark-parton model expectation.

The longitudinal spin-dependent structure function $g_1(x, Q^2)$ for deep-inelastic lepton-nucleon scattering has become increasingly important in unraveling the quark and gluon spin structure of the proton and neutron. The $g_1$ structure function depends both on $x$, the fractional momentum carried by the struck parton, and on $Q^2$, the four-momentum transfer squared of the virtual photons used as a probe of nucleon structure. Of particular interest are the fixed-$Q^2$ integrals $\Gamma_1^p(Q^2) = \int_0^1 g_1^p(x, Q^2)dx$ for the proton and $\Gamma_1^n(Q^2) = \int_0^1 g_1^n(x, Q^2)dx$ for the neutron. These integrals are directly related to the net quark helicity $\Delta q$ in the nucleon. Measurements of $\Gamma_1^p$ [1–5], $\Gamma_1^d$ [6–7], and $\Gamma_1^n$ [8] have found $\Delta q \approx 0.3$, significantly less than a prediction [9] that $\Delta q = 0.58$ assuming zero net strange quark helicity and SU(3) flavor symmetry in the baryon octet. A fundamental sum rule originally derived from current algebra by Bjorken [10] predicts the difference $\Gamma_1^p(Q^2) - \Gamma_1^n(Q^2)$. Recent measurements are in agreement with this sum rule prediction when perturbative QCD (pQCD) corrections [11] are included.

There are two main reasons for measuring $g_1$ over a wide range of $x$ and $Q^2$. The first is that experiments make measurements at fixed beam energies rather than at fixed $Q^2$. To evaluate first moment integrals of $g_1(x, Q^2)$ at constant $Q^2$ [typically between 2 and 10 (GeV/$c$)$^2$], extrapolations are needed. Data at low $x$ are at lower $Q^2$ than desired [as low as 1 (GeV/$c$)$^2$], while data at high $x$ are at higher $Q^2$ [up to 80 (GeV/$c$)$^2$]. Data at multiple beam energies allow for a measurement of the kinematic dependence of $g_1$, rather than relying on model-dependent extrapolations for the moment determinations.

The second motivation is that the kinematic dependence of $g_1$ can be used to obtain the underlying nucleon polarized quark and gluon distribution functions. According to the GLAP equations [12], $g_1$ is expected to evolve logarithmically with $Q^2$, increasing with $Q^2$ at low $x$, and decreasing with $Q^2$ at high $x$. A similar $Q^2$-dependence has been observed in the spin-averaged structure functions $F_1(x,Q^2)$ and $F_2(x,Q^2)$. For reference, in changing $Q^2$ from 2 to 10 (GeV/$c$)$^2$, $F_1$ decreases by 40% for $x \approx 0.5$, but increases by the same amount for $x \approx 0.035$ [13,14]. Since the GLAP equations are similar for $F_1$ and $g_1$, the $Q^2$ dependence of $g_1$ is expected to be similar to that of $F_1$, but the precise behavior is sensitive to the underlying spin-dependent quark and gluon distribution functions. Fits to polarized quark and gluon distribution functions have been made [15–19] using leading-order (LO) GLAP equations and data for $g_1(x,Q^2)$. Because of the limited $Q^2$ range and statistical precision of the data, constraints from QCD counting rules and Regge theory on the $x$-dependence have generally been imposed. Recently, fits have also been made [20,21] using next-to-leading-order (NLO) GLAP equations [17]. The results indicate that NLO fits are more sensitive to the strength of the polarized gluon distribution function $\Delta G(x,Q^2)$ than LO fits.

The theoretical interpretation of $g_1$ at low $Q^2$ is complicated by higher twist contributions not embodied in the GLAP equations. These terms are expected to be proportional to $C(x)/Q^2$, $D(x)/Q^4$, etc., where $C(x)$ and $D(x)$ are unknown functions. Higher twist contributions to the first moments $\Gamma_1^p$ and $\Gamma_1^n$ have been estimated to be only a few percent [22] for $Q^2 > 3$ (GeV/$c$)$^2$, but very little is known about their strength as a function of $x$.

In this Letter we study the $Q^2$ dependence of $g_1$ by supplementing our previously published results for $g_1^p$ [5], $g_1^d$ [7], and $g_2^p$ and $g_2^d$ [24] measured at average incident electron beam energy $E$ of 29.1 GeV with data for $g_1^p$ and $g_1^d$ at beam energies of 9.7 and 16.2 GeV. Data at all energies were taken at scattering angles of 4.5° and 7°. The ratio of polarized to unpolarized structure functions was determined from measured longitudinal asymmetries $A_\parallel$ using

$$g_1/F_1 = A_\parallel/d + (g_2/F_1)[(2Mx)/(2E-\nu)] \ , \qquad (1)$$

where $d = [(1-\epsilon)(2-y)]/\{y[1+\epsilon R(x,Q^2)]\}$, $y=\nu/E$, $\nu = E - E'$, $E'$ is the scattered electron energy, $\epsilon^{-1} = 1 + 2[1+\gamma^{-2}]\tan^2(\theta/2)$, $\gamma^2 = Q^2/\nu^2$, $\theta$ is the electron scattering angle, $M$ is the nucleon mass, and $R(x,Q^2) = [F_2(x,Q^2)(1+\gamma^2)]/[2xF_1(x,Q^2)] - 1$ is typically 0.2 for the kinematics of this experiment [14]. For the contribution of the transverse spin structure function $g_2$ we used the twist-two model of Wandzura and Wilczeck ($g_2^{WW}$) [23]

$$g_2(x, Q^2) = -g_1(x, Q^2) + \int_x^1 g_1(\xi, Q^2) d\xi/\xi \ , \tag{2}$$

evaluated with $g_1$ based on a global fit to the virtual photon asymmetry $A_1$ (see fits V, Table I). The $g_1$ and $g_2$ structure functions are related to $A_1$ (which is bounded by $|A_1| < 1$) by $A_1 = (g_1/F_1) - \gamma^2(g_2/F_1)$. The $g_2^{WW}$ model is in good agreement with our $g_2$ data at $E = 29$ GeV [24], the only energy at which both $A_{||}$ and the transverse asymmetry $A_\perp$ were measured. Using other reasonable models for $g_2$ (such as $g_2 = 0$) has relatively little impact on the results for $g_1$ due to the factor $2Mx/(2E - \nu)$ in Eq. 1.

The data analysis was essentially identical to that reported for the 29 GeV data [5,7], with $A_{||}$ calculated from the difference over the sum of rates for scattering longitudinally polarized electrons with spin either parallel or anti-parallel to polarized protons or deuterons in a cryogenic ammonia target. The most important corrections made were for the beam polarization (typically $0.85 \pm 0.02$), target polarization (typically $0.65 \pm 0.017$ for $NH_3$, $0.25 \pm 0.011$ for $ND_3$), fraction of polarizable nucleons (0.12 to 0.17 for $NH_3$, 0.22 to 0.24 for $ND_3$), and for contributions from polarized nitrogen atoms. Radiative corrections were calculated [25] using iterated global fits to all data (see fits V in Table I). The data at 29 GeV used here differ slightly from our previously published results [5,7] due to the new radiative corrections, the inclusion of more data runs, and improved measurements of the polarization of the target and beam. Data in the it resonance region defined by missing mass $W < 1.8$ GeV were not included in the present analysis, but those for $Q^2$ below the traditional deep-inelastic cutoff of $Q^2 = 1$ (GeV/$c$)$^2$ were kept.

The results for $g_1^p/F_1^p$ and $g_1^d/F_1^d$ are shown in Figs. 1 and 2, respectively, at eight values of $x$, and are listed in Table II. We display the ratio $g_1/F_1$ since it is closer to our measured asymmetries than $g_1$ alone, and because $g_1$ and $F_1$ are expected theoretically to have a similar $Q^2$ dependence, so that differences are emphasized in the ratio. Data from other experiments [1–4,6] are plotted using published longitudinal asymmetries $A_{||}$ and the same model for $R(x, Q^2)$ [14] and $g_2$ [23] as for the present data. Improved radiative corrections have been applied to the E80 [1] and E130 [2] results. Only statistical errors have been plotted. For the present experiment, most systematic errors (beam polarization, target polarization, fraction of polarizable nucleons in the target) for a given target are common to all data and correspond to an overall normalization error of about 5% for the proton data and 6% for the deuteron data. The remaining systematic errors (radiative corrections, model uncertainties for $R(x, Q^2)$, resolution corrections) vary smoothly with $x$ in a locally correlated fashion, ranging from a few percent for moderate $x$ bins, up to 15% for the highest and lowest $x$ bins at $E = 29$ GeV. For all data, the statistical errors dominate over the point-to-point systematic error.

The most striking feature of the data is that $g_1/F_1$ is approximately independent of $Q^2$ at fixed $x$, although there is a noticeable trend for the ratio to decrease for $Q^2 < 1$ (GeV/$c$)$^2$. To quantify the possible significance of this trend, we made two fits to the data. The first fit is motivated by possible differences in the twist-4 contributions to $g_1$ and $F_1$. We fit the data in each $x$ bin with the form $g_1/F_1 = a(1 + C/Q^2)$. The results for the $C$ coefficients are shown in Fig. 3 for all $Q^2$ [$Q^2 > 0.3$ (GeV/$c$)$^2$] (circles) and for $Q^2 > 1$ (GeV/$c$)$^2$ (squares). The coefficients indicate significantly negative values for $C$ at intermediate values of $x$ for the fits over all $Q^2$. The errors are much larger when data with $Q^2 < 1$ (GeV/$c$)$^2$ are excluded, and the resulting coefficients are consistent with no $Q^2$-dependence to $g_1/F_1$ ($C = 0$). There is no evidence for a significant $x$-dependence to $C$. Another fit to the data in each $x$ bin used the form $g_1/F_1 = a[1 + C\ln(1/Q^2)]$, motivated by looking for differences in the logarithmic evolution of $g_1$ and $F_1$. Again, the $C$ coefficients tend to be less than zero when no $Q^2$ cut is applied. The present data do not have sufficient precision to distinguish between a logarithmic and power law $Q^2$ dependence, but can rule out large differences between the $Q^2$-dependence of $g_1$ and $F_1$, especially for $Q^2 > 1$ (GeV/$c$)$^2$.

Shown in Figs. 1 and 2 as the dot-dashed curves are the low-$Q^2$ predictions from a representative global NLO pQCD fit [20] to all proton and deuteron data excluding those at the 9.7 GeV and 16.2 GeV beam energies of this experiment. This group [20] finds considerably less $Q^2$ dependence to $g_1/F_1$ when a *minimal* polarized gluon strength is used than when a *maximal* strength is chosen. Another group has made NLO pQCD fits to proton, deuteron, and neutron data using different constraints on the underlying parton distribution functions [21], examining the sensitivity to SU(3) symmetry breaking in the baryon $\beta$ decays. The results for their *standard* set are shown as the dotted curves in Figs. 1 and 2. Both [20] and [21] predict that $g_1^p/F_1^p$ increases with $Q^2$ in the moderate $x$ range ($0.03 < x < 0.3$), in agreement with the trend of our data when the $E = 9.7$ and $E = 16.2$ results (not included in their fits) are considered.

We also performed simple global fits to the data, both in order to have a practical parametrization (needed, for example, in making radiative corrections to the data), and to examine the possible effects of $Q^2$ dependence on the first moments $\Gamma_1$. Data points from SMC [4,6] at $x < 0.035$, not shown in Figs. 1 and 2, were included in the fits. The first fits are of the $Q^2$-independent form $g_1/F_1 = ax^\alpha(1 + bx + cx^2)$, with the constraint that $A_1 = g_1/F_1 - \gamma^2 g_2^{WW}/F_1 \to 1$ for $x \to 1$ at $Q^2 = 2$ (GeV/$c$)$^2$. As can be seen in Table I (case I), the fits to all the proton and deuteron data are acceptable (combined $\chi^2 = 125$ for 104 d.f.), but the fits systematically lie above the lowest $Q^2$ points. The fits are improved ($\chi^2 = 94$ for 82 d.f.) by excluding the data for $Q^2 < 1$ (GeV/$c$)$^2$ (case II in Table I and dashed curves in Figs. 1 and 2). Better fits are obtained by introducing an overall multiplicative correction term of the form $(1 + C/Q^2)$ to account for the low $Q^2$ data ($\chi^2 = 104$ for 102 d.f.), as shown by the solid lines in Figs. 1 and 2 (case III in Tabl I). Using an $x$-independent value of $C$ is reasonable given the results shown in Fig. 3. We examined an alternate correction term of the form $[1 + C\ln(1/Q^2)]$ (case IV in Table I) which shows an intermediate level of improvement ($\chi^2 = 113$ for 102 d.f.). We also examined the $Q^2$-dependence of $A_1$, extracted from measured values of $A_{||}$ and using the $g_2^{WW}$ model for $g_2$. The $x$ coefficients listed in Table I (case V) are somewhat different from the $g_1/F_1$ fits, but the $C$ coefficients remain negative. Thus both $A_1$ and $g_1/F_1$ indicate a significant tendency to decrease at low $Q^2$ when the low $Q^2$ data are included in the fits.

We have evaluated the first moments $\Gamma_1^p$ and $\Gamma_1^d$, and the corresponding results for $\Gamma_1^p - \Gamma_1^n$, using the $Q^2$-independent fits II ($Q^2 > 1$ (GeV/$c$)$^2$) and the $Q^2$-dependent fits III (all $Q^2$) shown in Table I. A global fit [13,14] was used for $F_1$ to obtain $g_1$ from $g_1/F_1$. The results for $\Gamma_1^p - \Gamma_1^n$ are shown as a function of $Q^2$ as the lower (fit II) and upper (fit III) bands in Fig. 4, where the width of the band reflects the combined statistical and systematic error estimate. Both fits are in reasonable agreement with the Bjorken sum rule, shown as the solid curve, evaluated using $\alpha_s(Q^2)$ evolved in $Q^2$ from $\alpha_s(M_Z) = 0.117 \pm 0.005$ [26] for the QCD corrections [11] taken to third order in $\alpha_s$. Alternatively, if we assume the sum rule is correct, we can use the measured $\Gamma_1^p(Q^2) - \Gamma_1^n(Q^2)$ to determine the strong coupling $\alpha_s$. The case II ($Q^2$-independent $g_1/F_1$) fits to the proton and deuteron data integrated at $Q^2 = 3$ (GeV/$c^2$) yield $\alpha_s(M_Z) = 0.119^{+0.007}_{-0.019}$, while the case III ($Q^2$-dependent $g_1/F_1$) fits yield $\alpha_s(M_Z) = 0.113^{+0.011}_{-0.035}$, both in agreement with the world average result of 0.117.

We have examined the sensitivity to the possible $Q^2$ dependence of $g_1/F_1$ of the net quark helicity $\Delta q$ extracted from global fits to the data. We computed $\Delta q$ using [27]

$$\Delta q = \frac{9}{c_s(Q^2)}\Big[\Gamma_1^p(Q^2) - \Big(\frac{F+D}{12} + \frac{3F-D}{36}\Big)c_{ns}(Q^2)\Big] , \tag{3}$$

with $F+D = 1.2573 \pm 0.0028$ [26], $F/D = 0.575 \pm 0.016$ [27], extracted assuming SU(3) flavor symmetry in the baryon octet. The singlet and non-singlet QCD correction factors $c_s(Q^2)$ and $c_{ns}(Q^2)$ are given in [11,28]. At $Q^2 = 3$ (GeV/$c$)$^2$, we obtain $\Delta q = 0.34 \pm 0.09$ for global proton fit II, and $\Delta q = 0.36 \pm 0.10$ for proton fit III, somewhat higher than $\Delta q = 0.27 \pm 0.10$, obtained using the previous analysis of the E143 $E = 29$ GeV data only [5], which assumed $g_1/F_1$ independent of $Q^2$. For the deuteron fits, we used

$$\Delta q = \frac{9}{c_s(Q^2)}\Big[\frac{\Gamma_1^d(Q^2)}{1 - 1.5\omega_d} - \Big(\frac{3F-D}{36}\Big)c_{ns}(Q^2)\Big] , \tag{4}$$

where $\omega_d$ is the $D$-state probability in the deuteron, to obtain $\Delta q = 0.35 \pm 0.05$ for fit II, and $\Delta q = 0.34 \pm 0.05$ for fit III, again somewhat higher than our previous deuteron analysis $\Delta q = 0.30 \pm 0.06$ [7], but in good agreement with the new proton results. For both targets, using the $Q^2$-independent fit II or the $Q^2$-dependent fit III makes little difference at $Q^2 = 3$ (GeV/$c$)$^2$, but we find $\Delta q$ (which should be independent of $Q^2$) to vary less with $Q^2$ for fit III than for fit II, especially for the deuteron fits.

In summary, the assumption that $g_1$ and $F_1$ have approximately the same $Q^2$-dependence has been found to be consistent with all available data in the deep inelastic region $Q^2 > 1$ (GeV/$c$)$^2$, although significant deviations from this assumption are found at lower $Q^2$. Global fits to the data with and without a possible $Q^2$ dependence to $g_1/F_1$ provide a useful parametrization of available data, and validate previous conclusions that the fundamental Bjorken sum rule is satisfied, and that the net quark helicity content of the nucleon is less than expected in the simple relativistic parton model.

We thank the authors of Refs. [18–21] for valuable discussions and for sending numerical results of their calculations.

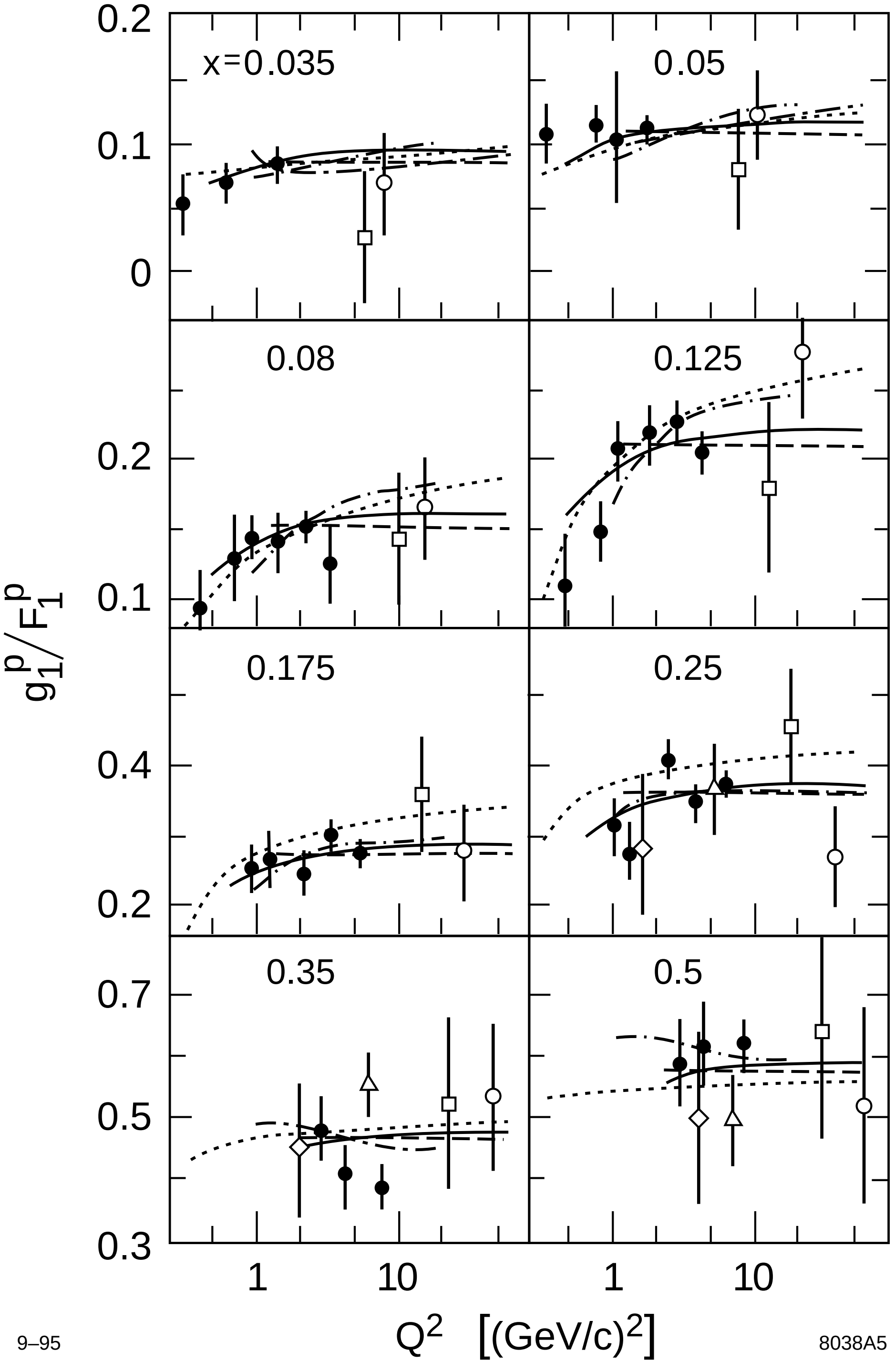

Figure 1. Ratios $g_1^p/F_1^p$ extracted from experiments assuming the $g_2^{WW}$ model for $g_2$. The errors are statistical only. Data are from this experiment (solid circles), SLAC E80 [1] (diamonds), SLAC E130 [2] (triangles), EMC [3] and SMC [4] (open circles). The dashed and solid curves correspond to global fits II ($g_1^p/F_1^p$ $Q^2$-independent) and III ($g_1^p/F_1^p$ $Q^2$-dependent) in Table I, respectively. Representative NLO pQCD fits from Refs. [20] [21] are shown as the dot-dashed and dotted curves, respectively.

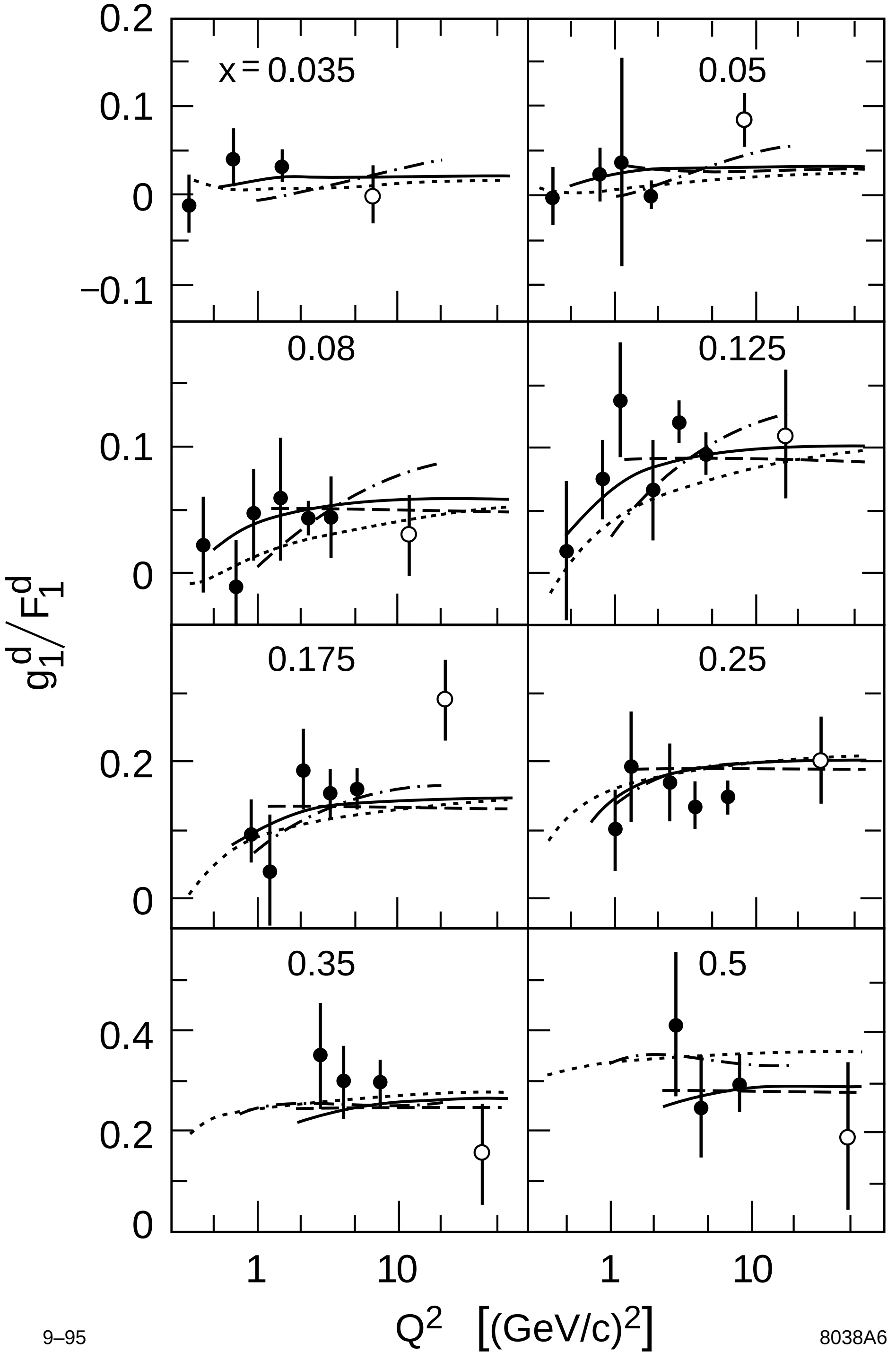

Figure 2. Ratios $g_1^d/F_1^d$ from this experiment (solid circles) and SMC [6] (open circles). The curves are as in Fig. 1.

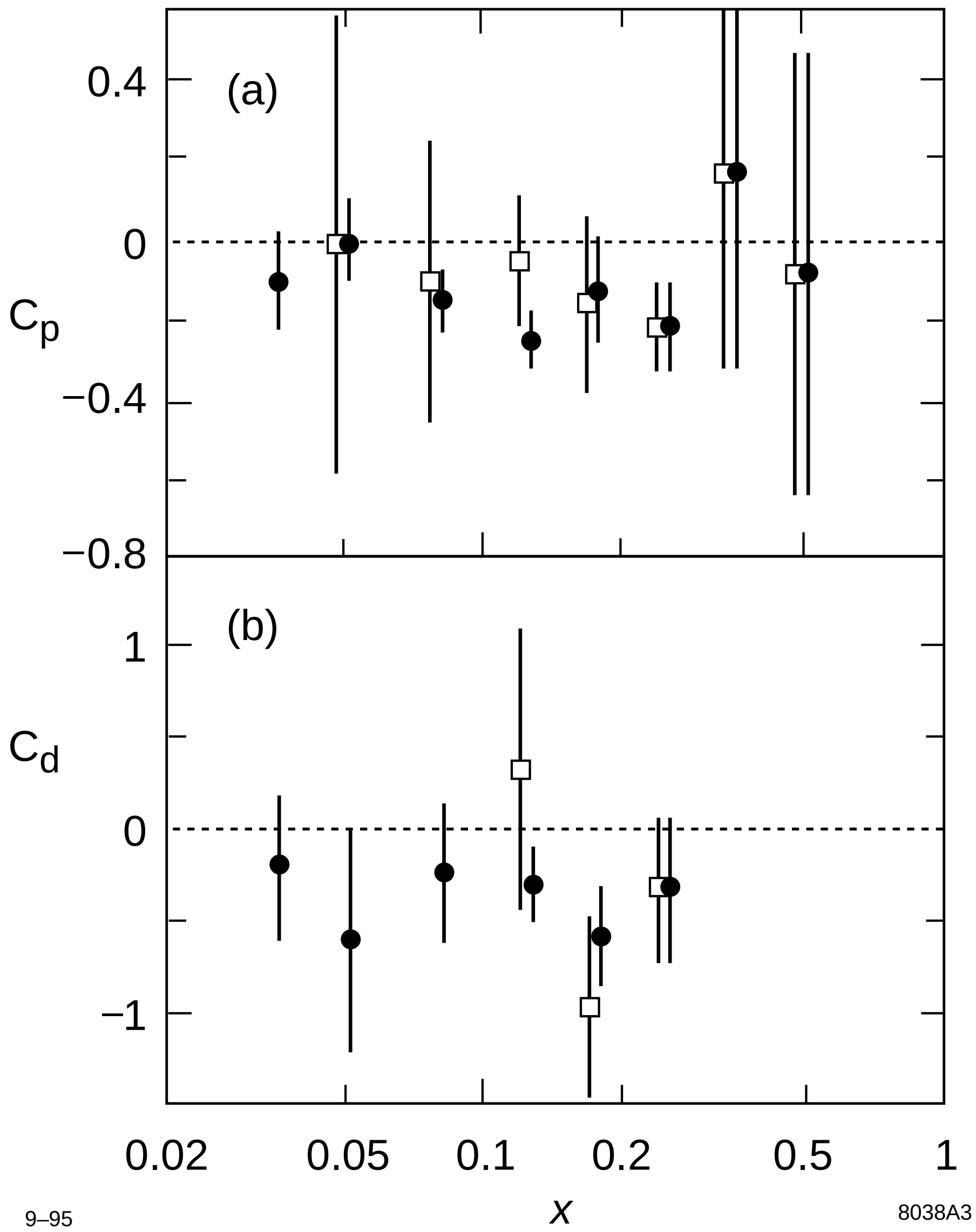

Figure 3. Coefficients $C$ for fits to $g_1/F_1$ at fixed $x$ of the form $a(1+C/Q^2)$ for (a) proton and (b) deuteron. Solid circles are fits to all data [$Q^2 > 0.3$ (GeV/$c$)$^2$], and open squares are fits only to data with $Q^2 > 1$ (GeV/$c$)$^2$.

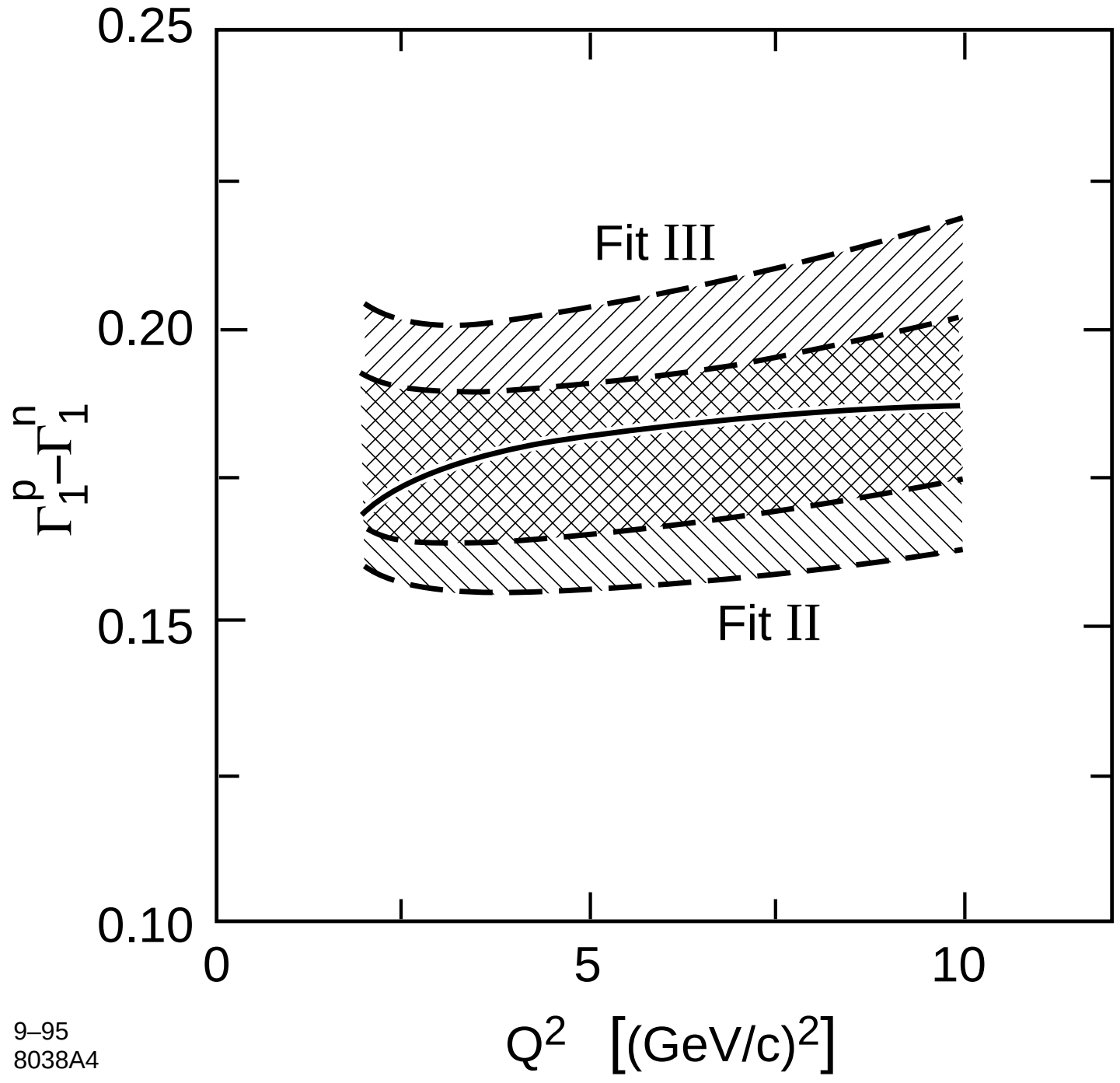

Figure 4. Evaluations of $\Gamma_1^p - \Gamma_1^n$ from the $Q^2$-independent fits II (lower band) and $Q^2$-dependent fits III (upper band) listed in Table I. The errors include both statistical and systematic contributions and are indicated by the widths of the bands. The solid curve is the prediction of the Bjorken sum rule with third-order QCD corrections.

**TABLE I. Coefficients for fits to all available data with $Q^2 > Q^2_{min}$ of the form $ax^{\alpha}(1 + bx + cx^2)[1 + Cf(Q^2)]$, along with the $\chi^2$ for the indicated number of degrees of freedom, calculated with statistical errors only. Fits to IV are to $g_1/F_1$, while fit V is to $A_1$.**

| fit to | $Q^2_{min}$ | $f(Q^2)$ | $\alpha$ | $a$ | $b$ | $c$ | $C$ | $\chi^2$ | d.f. |
|---|---|---|---|---|---|---|---|---|---|
| I. $g_1^p/F_1^p$ | 0.3 | none | 0.50 | 0.380 | 4.767 | -4.812 | 0 | 64 | 59 |
| II. $g_1^p/F_1^p$ | 1.0 | none | 0.56 | 0.513 | 2.948 | -3.242 | 0 | 40 | 48 |
| III. $g_1^p/F_1^p$ | 0.3 | $1/Q^2$ | 0.50 | 0.455 | 3.533 | -3.677 | -0.140 | 48 | 58 |
| IV. $g_1^p/F_1^p$ | 0.3 | $\ln(1/Q^2)$ | 0.56 | 0.487 | 2.422 | -2.717 | -0.080 | 55 | 58 |
| V. $A_1^p$ | 0.3 | $1/Q^2$ | 0.56 | 0.590 | 1.871 | -1.028 | -0.160 | 51 | 58 |
| I. $g_1^d/F_1^d$ | 0.3 | none | 1.54 | 2.760 | -1.941 | 1.072 | 0 | 61 | 45 |
| II. $g_1^d/F_1^d$ | 1.0 | none | 1.48 | 2.532 | -1.908 | 1.051 | 0 | 54 | 34 |
| III. $g_1^d/F_1^d$ | 0.3 | $1/Q^2$ | 1.44 | 2.612 | -1.946 | 1.109 | -0.300 | 56 | 44 |
| IV. $g_1^d/F_1^d$ | 0.3 | $\ln(1/Q^2)$ | 1.46 | 2.063 | -2.015 | 1.175 | -0.140 | 58 | 44 |
| V. $A_1^d$ | 0.3 | $1/Q^2$ | 1.46 | 2.802 | -2.125 | 1.549 | -0.320 | 56 | 44 |

TABLE II. Results for $g_1^p/F_1^p$ and $g_1^d/F_1^d$ from this experiment, extracted assuming the $g_2^{WW}$ model for $g_2$. Both statistical and total systematic errors are listed. The boundaries between the $x$ bins are at 0.03, 0.04, 0.06, 0.1, 0.15, 0.2, 0.3, 0.4, and 0.6.

| $x$ | $\langle Q^2\rangle$ (GeV/$c$)$^2$ | $E$ (GeV) | $(g_1^p/F_1^p) \pm$ stat $\pm$ syst | $(g_1^d/F_1^d) \pm$ stat $\pm$ syst |
|---|---|---|---|---|
| 0.035 | 0.32 | 9.7 | $0.053 \pm 0.022 \pm 0.021$ | $-0.020 \pm 0.031 \pm 0.009$ |
| 0.035 | 0.65 | 16.2 | $0.069 \pm 0.014 \pm 0.011$ | $0.039 \pm 0.035 \pm 0.030$ |
| 0.035 | 1.45 | 29.1 | $0.082 \pm 0.014 \pm 0.008$ | $0.033 \pm 0.015 \pm 0.011$ |
| 0.050 | 0.37 | 9.7 | $0.110 \pm 0.023 \pm 0.024$ | $0.004 \pm 0.033 \pm 0.010$ |
| 0.050 | 0.79 | 16.2 | $0.117 \pm 0.014 \pm 0.013$ | $0.023 \pm 0.030 \pm 0.017$ |
| 0.050 | 1.14 | 16.2 | $0.107 \pm 0.051 \pm 0.013$ | $0.038 \pm 0.116 \pm 0.014$ |
| 0.050 | 1.82 | 29.1 | $0.113 \pm 0.011 \pm 0.008$ | $-0.001 \pm 0.012 \pm 0.008$ |
| 0.080 | 0.42 | 9.7 | $0.095 \pm 0.026 \pm 0.027$ | $0.031 \pm 0.038 \pm 0.012$ |
| 0.080 | 0.71 | 9.7 | $0.129 \pm 0.029 \pm 0.024$ | $-0.010 \pm 0.042 \pm 0.010$ |
| 0.080 | 0.95 | 16.2 | $0.144 \pm 0.015 \pm 0.016$ | $0.048 \pm 0.034 \pm 0.012$ |
| 0.080 | 1.48 | 16.2 | $0.140 \pm 0.020 \pm 0.014$ | $0.059 \pm 0.047 \pm 0.011$ |
| 0.080 | 2.33 | 29.1 | $0.150 \pm 0.010 \pm 0.011$ | $0.044 \pm 0.012 \pm 0.006$ |
| 0.080 | 3.38 | 29.1 | $0.131 \pm 0.028 \pm 0.011$ | $0.039 \pm 0.031 \pm 0.007$ |
| 0.125 | 0.47 | 9.7 | $0.110 \pm 0.037 \pm 0.031$ | $0.022 \pm 0.055 \pm 0.016$ |
| 0.125 | 0.85 | 9.7 | $0.150 \pm 0.020 \pm 0.025$ | $0.073 \pm 0.030 \pm 0.012$ |
| 0.125 | 1.13 | 16.2 | $0.209 \pm 0.022 \pm 0.019$ | $0.138 \pm 0.044 \pm 0.013$ |
| 0.125 | 1.90 | 16.2 | $0.221 \pm 0.019 \pm 0.015$ | $0.066 \pm 0.039 \pm 0.009$ |
| 0.125 | 2.94 | 29.1 | $0.227 \pm 0.014 \pm 0.015$ | $0.121 \pm 0.017 \pm 0.007$ |
| 0.125 | 4.42 | 29.1 | $0.203 \pm 0.014 \pm 0.013$ | $0.095 \pm 0.017 \pm 0.007$ |
| 0.175 | 0.95 | 9.7 | $0.254 \pm 0.032 \pm 0.026$ | $0.107 \pm 0.047 \pm 0.014$ |
| 0.175 | 1.24 | 16.2 | $0.265 \pm 0.040 \pm 0.024$ | $0.040 \pm 0.081 \pm 0.017$ |
| 0.175 | 2.20 | 16.2 | $0.244 \pm 0.029 \pm 0.019$ | $0.189 \pm 0.059 \pm 0.012$ |
| 0.175 | 3.37 | 29.1 | $0.297 \pm 0.025 \pm 0.018$ | $0.155 \pm 0.031 \pm 0.011$ |
| 0.175 | 5.33 | 29.1 | $0.270 \pm 0.019 \pm 0.016$ | $0.165 \pm 0.023 \pm 0.009$ |
| 0.250 | 1.02 | 9.7 | $0.315 \pm 0.038 \pm 0.027$ | $0.105 \pm 0.058 \pm 0.015$ |
| 0.250 | 1.33 | 16.2 | $0.281 \pm 0.039 \pm 0.031$ | $0.196 \pm 0.080 \pm 0.025$ |
| 0.250 | 2.52 | 16.2 | $0.411 \pm 0.026 \pm 0.024$ | $0.173 \pm 0.054 \pm 0.017$ |
| 0.250 | 3.77 | 29.1 | $0.348 \pm 0.025 \pm 0.022$ | $0.138 \pm 0.032 \pm 0.015$ |
| 0.250 | 6.42 | 29.1 | $0.373 \pm 0.017 \pm 0.021$ | $0.151 \pm 0.021 \pm 0.013$ |
| 0.350 | 2.80 | 16.2 | $0.480 \pm 0.050 \pm 0.029$ | $0.350 \pm 0.104 \pm 0.023$ |
| 0.350 | 4.14 | 29.1 | $0.405 \pm 0.054 \pm 0.027$ | $0.300 \pm 0.072 \pm 0.020$ |
| 0.350 | 7.50 | 29.1 | $0.391 \pm 0.033 \pm 0.027$ | $0.298 \pm 0.042 \pm 0.018$ |
| 0.500 | 2.97 | 16.2 | $0.590 \pm 0.070 \pm 0.033$ | $0.411 \pm 0.141 \pm 0.028$ |
| 0.500 | 4.38 | 29.1 | $0.617 \pm 0.069 \pm 0.034$ | $0.246 \pm 0.094 \pm 0.025$ |
| 0.500 | 8.36 | 29.1 | $0.629 \pm 0.038 \pm 0.034$ | $0.293 \pm 0.051 \pm 0.022$ |